\documentclass[10pt, a4paper]{article}

\usepackage{lrec-coling2024}

\usepackage{color}
\usepackage[noend]{algpseudocode}
\usepackage{bm}
\usepackage{algorithmicx,algorithm}
\usepackage{multirow}
\usepackage{subcaption}
\usepackage{amsfonts}
\usepackage{amsmath}
\usepackage{booktabs}
\usepackage{makecell}

\title{ToolRerank: Adaptive and Hierarchy-Aware Reranking \\ for Tool Retrieval}

\name{Yuanhang Zheng$^1$, Peng Li$^{2,3*}$, Wei Liu$^{4*}$, Yang Liu$^{1,2,3}$, Jian Luan$^4$, Bin Wang$^4$\thanks{$^*$\ Corresponding authors: Peng Li and Wei Liu}} 

\address{
$^1$Department of Computer Science and Technology, Tsinghua University, Beijing, China \\
$^2$Institute for AI Industry Research (AIR), Tsinghua University, Beijing, China \\
$^3$Shanghai Artificial Intelligence Laboratory, Shanghai, China \\
$^4$Xiaomi AI Lab \\
zheng-yh19@mails.tsinghua.edu.cn, lipeng@air.tsinghua.edu.cn, liuyang2011@tsinghua.edu.cn \\
\{liuwei40, luanjian, wangbin11\}@xiaomi.com \\
}

\abstract{
Tool learning aims to extend the capabilities of large language models (LLMs) with external tools.
A major challenge in tool learning is how to support a large number of tools, including unseen tools.
To address this challenge, previous studies have proposed retrieving suitable tools for the LLM based on the user query.
However, previously proposed methods do not consider the differences between seen and unseen tools, nor do they take the hierarchy of the tool library into account, which may lead to suboptimal performance for tool retrieval.
Therefore, to address the aforementioned issues, we propose ToolRerank, an adaptive and hierarchy-aware reranking method for tool retrieval to further refine the retrieval results.
Specifically, our proposed ToolRerank includes Adaptive Truncation, which truncates the retrieval results related to seen and unseen tools at different positions, and Hierarchy-Aware Reranking, which makes retrieval results more concentrated for single-tool queries and more diverse for multi-tool queries.
Experimental results show that ToolRerank can improve the quality of the retrieval results, leading to better execution results generated by the LLM.
 \\ \newline  }

\begin{document}

\maketitleabstract

\section{Introduction}

Recently, large language models (LLMs) have achieved impressive performance on various tasks~\citep{openaichatgptblog,openai2023gpt4}. However, LLMs may still struggle to solve certain types of problems effectively. For example, LLMs are usually incapable of answering questions about the latest events without additional assistance~\citep{openaichatgptblog}. Moreover, mathematical problems and low-resource languages can also pose challenges for LLMs~\citep{patel2021nlp,lin2022fewshot}. Thus, to extend the capabilities of the LLMs, various studies have proposed tool learning, which augments LLMs with external tools~\citep{schick2023toolformer,chen2023chatcot,yao2023react}.

A vanilla method for tool learning is to provide API documents in the input context of the LLMs~\citep{shen2023hugginggpt,hsieh2023tool}.
However, the number of provided documents is limited by the maximum context length of the LLMs, making it difficult to use this method when a large number of tools are available~\citep{qin2023tool}.
To address this challenge, one possible solution is to fine-tune the LLMs to enable them to use tools without provided documents~\citep{hao2023toolkengpt}.
However, this method is inconvenient for supporting new tools, as the model needs to be fine-tuned again.
In contrast, retrieval-based methods, which retrieve suitable tools from a tool library, can be generalized to unseen tools without further training~\citep{paranjape2023art,patil2023gorilla,qin2023toolllm}.
Thus, we mainly focus on retrieval-based methods in this work.

\begin{figure*}[t]
    \centering
    \begin{subfigure}[b]{0.27\textwidth}
        \centering
        \includegraphics[scale=0.6]{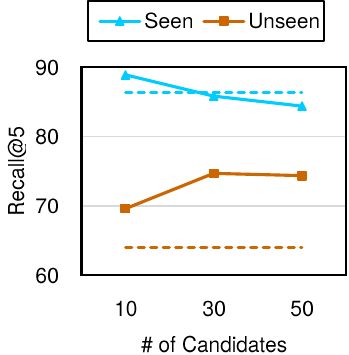} 
        \caption{}
    \end{subfigure}
    \begin{subfigure}[b]{0.72\textwidth}
        \centering
        \includegraphics[scale=0.6]{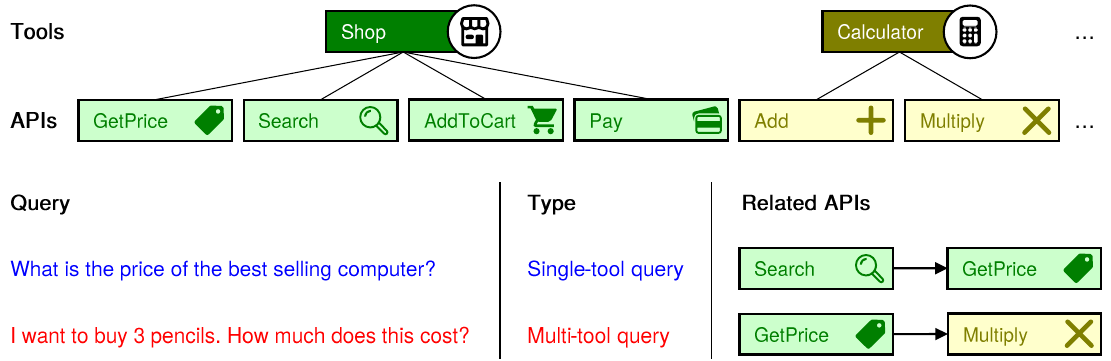} 
        \caption{}
    \end{subfigure}
    \caption{(a) Recall@5 of the reranked retrieval results for seen and unseen tools with different number of candidates given to the reranker. The dashed lines represent the retrieval performance without using the reranker. (b) Example of the hierarchy of the tool library, a single-tool query and a multi-tool query.}
    \label{fig:recall_hierarchy}
\end{figure*}

Existing retrieval-based methods mainly differ in the retrievers being used, which can be roughly categorized into three types: BM25-based, LLM-based, and dual-encoder-based. However, these methods still have some limitations. Specifically, BM25 retrievers rely on literal similarity~\citep{robertson2009probabilistic} and thus they usually cannot capture semantic relations, resulting in an inferior performance~\citep{qin2023toolllm}.
LLM-based retrievers use the LLM to assess the suitability of the tool for the query~\citep{paranjape2023art} and thus their efficiency is limited by the inference speed of the LLM.
Dual-encoder-based retrievers choose suitable tools based on the cosine similarity between the query and documents, which are computed using two independent encoders~\citep{karpukhin2020dense}.
Although they strike a balance between effectiveness and efficiency, they lack fine-grained interaction between the query and the documents~\cite{humeau2020poly} and their performance is still far from perfect for tool learning~\citep{qin2023toolllm}.
Therefore, how to retrieve tools both efficiently and effectively still remains a challenge.

In the research field of information retrieval, reranking is a widely used technique to enhance the retrieval performance, where a more computationally intensive but effective model is used to refine the retrieval results~\citep{guo2016deep,xiong2017endtoend,nogueira2019passage,yan2019idst}.
Typically, we may rerank a fixed number of candidates using a cross-encoder reranker, allowing for fine-grained interaction between the query and the document~\cite{nogueira2019passage}.
However, there exist some issues when applying such method to tool retrieval.
On the one hand, as shown in Figure~\ref{fig:recall_hierarchy}(a), the reranker behaves differently for seen and unseen tools. Specifically, the reranker performs better with fewer candidates for seen tools and with more candidates for unseen tools. This indicates that a fixed number of candidates may result in suboptimal performance for tool retrieval.
On the other hand, as shown in Figure~\ref{fig:recall_hierarchy}(b), the tool library may have a hierarchy where a tool may have multiple APIs~\citep{qin2023toolllm}. In reality, some queries should be resolved using different APIs of a single tool (single-tool queries). In such cases, it is ideal for the retrieved APIs to belong to a single tool. Other queries should be resolved using APIs of different tools (multi-tool queries). For these queries, it is preferable for the retrieved APIs to belong to diverse tools.
Unfortunately, the reranker itself is unable to take these issues into account, leading to suboptimal performance for tool retrieval.

Therefore, to address the aformentioned issues, we propose ToolRerank, an adaptive and hierarchy-aware reranking method for tool retrieval. Specifically, ToolRerank includes two key components: Adaptive Truncation and Hierarchy-Aware Reranking.
First, to adapt to the behavior of the reranker, we propose Adaptive Truncation, which truncates the results related to seen and unseen tools at different positions.
Second, to take advantage of the hierarchy of the tool library, we propose Hierarchy-Aware Reranking to further rerank the retrieval results, making the fine-grained retrieval results more concentrated for single-tool queries and more diverse for multi-tool queries.
Experimental results on the ToolBench~\citep{qin2023toolllm} dataset show that ToolRerank can improve the quality of the retrieval results, leading to better execution results generated by the LLM. The code is available at \url{https://github.com/XiaoMi/ToolRerank}.

\section{Preliminaries}

\subsection{Tool Retrieval}

Tool retrieval aims to retrieve the most suitable APIs from a tool library based on a given user query~\citep{paranjape2023art,patil2023gorilla,qin2023toolllm}. Generally, the tool library may have a hierarchy where a tool may have multiple APIs, and the retrieved APIs may belong to one or more tools. Then, to enable the LLM to call the APIs, we give the documents of the retrieved APIs to the LLM as the input context~\citep{qin2023tool}.

Formally, we use a retriever to choose $k$ most suitable APIs $C=[c_1,\dots,c_k]$ from the tool library based on the user query $q$. For each $c_i$ in $C$, we use $\mathrm{tool}(c_i)$ to represent the tool which $c_i$ belongs to. Then, we add the document of the retrieved APIs to the input context, and use the LLM to call the APIs and generate the execution result:
\begin{equation}
y=\mathrm{LLM}(p_s,c_1,\dots,c_k,q),
\end{equation}

\noindent where $y$ denotes the output of the LLM, $\mathrm{LLM}(\cdot,\dots,\cdot)$ denotes the inference function of the LLM, and $p_s$ denotes the system prompt.

\begin{figure*}[t]
\begin{center}
\includegraphics[scale=0.6]{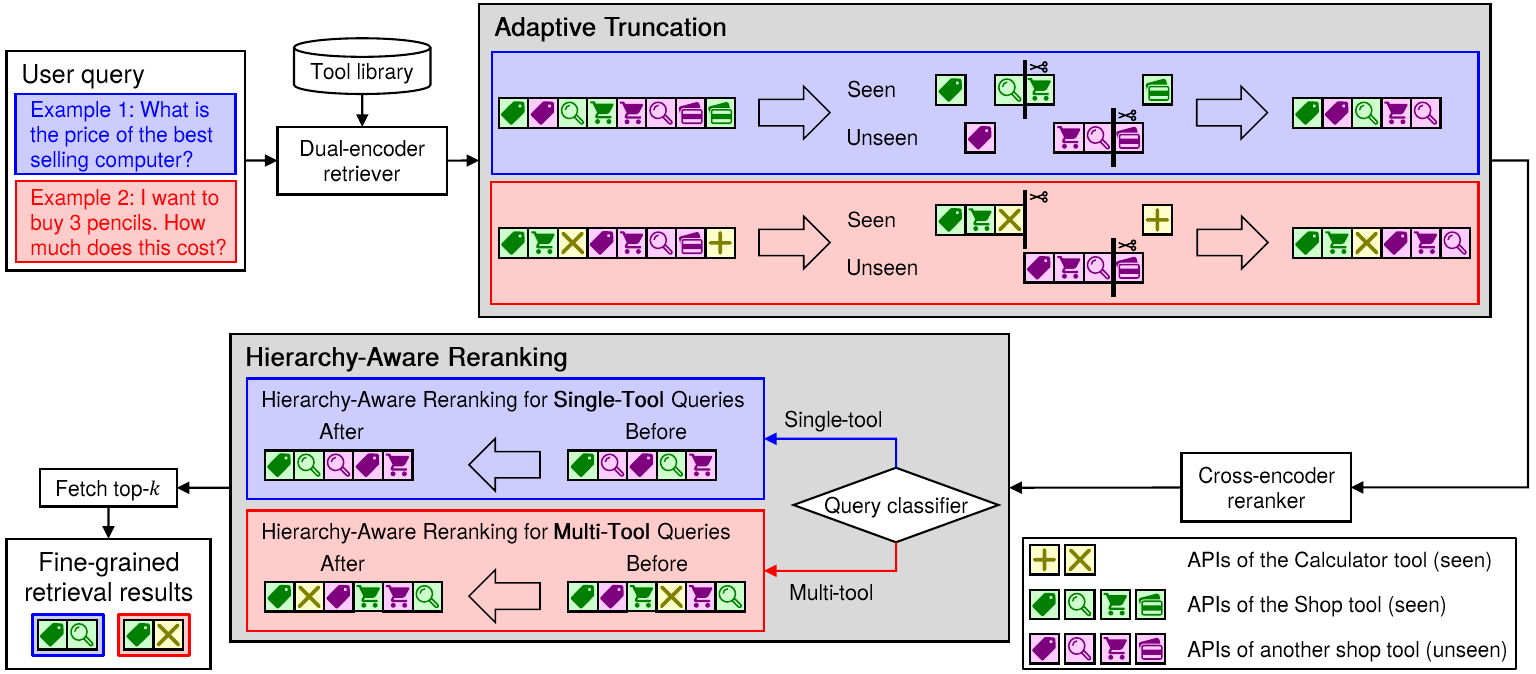} 
\caption{Overview of our proposed ToolRerank. We use Adaptive Truncation (Section~\ref{sec:hybrid_truncation}) to truncate the coarse-grained retrieval results related to seen and unseen tools at different positions. We use Hierarchy-Aware Reranking (Section~\ref{sec:separate_algorithms}) to further rerank the results to make the fine-grained retrieval results more concentrated for single-tool queries and more diverse for multi-tool queries. The execution processes only related to \textcolor{blue}{Example 1} or \textcolor{red}{Example 2} are marked in \textcolor{blue}{blue} and \textcolor{red}{red}, respectively.}
\label{fig:flowchart}
\end{center}
\end{figure*}

\subsection{Dual-Encoder Retriever}

The dual-encoder retriever~\citep{karpukhin2020dense} encodes user queries and documents into dense vectors, and retrieves the relevant documents based on the cosine similarity between the queries and documents.

Formally, the dual-encoder retriever $f_{\mathrm{dual}}(\cdot)$ encodes a user query $q$ or a document $d$ into a dense vector $\mathbf{v}_q$ or $\mathbf{v}_d$:
\begin{align}
\mathbf{v}_q=f_{\mathrm{dual}}(q), \\
\mathbf{v}_d=f_{\mathrm{dual}}(d).
\end{align}

Then, we calculate the cosine similarity between $\mathbf{v}_q$ and $\mathbf{v}_d$:
\begin{equation}
\mathrm{sim}(q,d)=\frac{\mathbf{v}_q \cdot \mathbf{v}_d}{||\mathbf{v}_q||\times||\mathbf{v}_d||}.
\end{equation}

Finally, we return the coarse-grained retrieval results by choosing $m$ different documents with the highest cosine similarity.

\subsection{Cross-Encoder Reranker}

The cross-encoder reranker~\citep{nogueira2019passage} takes the concatenation of the user query and the document as the input and outputs a relevance score between 0 and 1 which denotes whether the document is relevant to the query.

Formally, given a user query $q$ and a document $d$, the cross-encoder reranker $f_{\mathrm{cross}}(\cdot,\cdot)$ calculates the relevance score $\mathrm{score}(q,d)$:
\begin{equation}
\mathrm{score}(q,d)=f_{\mathrm{cross}}(q,d),
\end{equation}

Then, we refine the coarse-grained retrieval results and obtain the fine-grained retrieval results by choosing $k$ different documents with the highest relevance scores.

\section{Methodology}

In this section, we introduce ToolRerank in detail. First, in Section~\ref{sec:overall_procedure}, we describe the overall procedure of ToolRerank. Then, in Section~\ref{sec:hybrid_truncation}, we introduce Adaptive Truncation, which truncates the retrieval results related to seen and unseen tools at different positions. Finally, in Section~\ref{sec:separate_algorithms}, we describe Hierarchy-Aware Reranking, which makes the retrieval results more concentrated for single-tool queries and diverse for multi-tool queries.

\subsection{Overall Procedure}
\label{sec:overall_procedure}

As illustrated in Figure~\ref{fig:flowchart}, ToolRerank returns fine-grained retrieval results with $k$ different APIs based on the user query. First, we use a dual-encoder retriever~\citep{karpukhin2020dense} to obtain coarse-grained retrieval results $C=[c_1,\dots,c_m]$. Then, we apply Adaptive Truncation to $C$ to obtain the truncated results $T=[t_1, \dots, t_l]$. Next, we use a cross-encoder reranker~\citep{nogueira2019passage} to rerank $T$ and obtain the reranked results $R=[r_1, \dots, r_l]$.

To take the hierarchy of the tool library into account, we use Hierarchy-Aware Reranking to further rerank the retrieval results. To achieve this, we use a classifier to distinguish the single- and multi-tool user queries. Then, based on the classification result, we further rerank $R$ using different reranking algorithms to obtain the final reranked results $F=[f_1,\dots,f_l]$. Finally, we fetch the top-$k$ results in $F$ to obtain the fine-grained retrieval results and give them to the LLM.

\subsection{Adaptive Truncation}
\label{sec:hybrid_truncation}

As demonstrated in Figure~\ref{fig:recall_hierarchy}(a), the number of provided candidates may affect the performance of the reranker. Specifically, if the user query is related to unseen tools, increasing the number of candidates improves the retrieval performance. Otherwise, giving more candidates will decrease the performance. Thus, to better adapt to both seen and unseen tools, we propose Adaptive Truncation, which truncates the retrieval results related to seen and unseen tools at different positions.

Formally, we set two different thresholds $m_s$ and $m_u$ ($m_s<m_u$) for truncating the results. Then, for each API $c_i$ in $C$, if $\mathrm{tool}(c_i)$ is seen in the training data, we add $c_i$ to $T$ when its position $i$ satisfies $i \leq m_s$. Otherwise, we add $c_i$ to $T$ when $i \leq m_u$.

\begin{algorithm}[t]
\caption{Hierarchy-Aware Reranking for Single-Tool Queries \label{alg:single_tool}}
{\bf Input:}
user query $q$, reranked results $R=[r_1, \dots, r_l]$ \\
{\bf Output:}
final reranked results $F=[f_1, \dots, f_l]$
\begin{algorithmic}[1]
\State $\mathcal{X}\leftarrow\{\mathrm{tool}(r_1)\}$
\For {$i\leftarrow 2$ to $l$}
    \If {$\mathrm{score}(q, r_i) > {\tau}_s$}
        \State $\mathcal{X}\leftarrow\mathcal{X}\cup\{\mathrm{tool}(r_i)\}$
    \EndIf
\EndFor
\State $F_1\leftarrow F_2 \leftarrow [\ ]$
\For {$i\leftarrow 1$ to $l$}
    \If {$\mathrm{tool}(r_i)\in\mathcal{X}$}
        \State $F_1\leftarrow F_1+[r_i]$
    \Else
        \State $F_2\leftarrow F_2+[r_i]$
    \EndIf
\EndFor
\State $F\leftarrow F_1+F_2$
\State \Return $F$
\end{algorithmic}
\end{algorithm}

\subsection{Hierarchy-Aware Reranking}
\label{sec:separate_algorithms}

As shown in Figure~\ref{fig:recall_hierarchy}(b), according to the hierarchy of the tool library, the user queries can be divided into two categories: single-tool queries and multi-tool queries~\citep{qin2023toolllm}. 
Thus, we propose Hierarchy-Aware Reranking to address these two types of queries more effectively. First, to distinguish these two types of queries, we introduce a classifier which is trained on the training data of the retriever. Then, based on the classification results, we use different reranking algorithms to further rerank the retrieval results to make the fine-grained retrieval results more concentrated for single-tool queries and diverse for multi-tool queries.

\paragraph{Hierarchy-Aware Reranking for single-tool queries.} For single-tool queries, it is ideal for the fine-grained results to belong to a single tool. However, the positive results recognized by the cross-encoder reranker may belong to multiple tools, since there exist some functionally similar tools in the tool library and the reranker may fail to determine which is the correct tool for resolving the user query. To strike a balance between these two considerations, we consider that a tool may be suitable for resolving the query if it contains at least one API that the reranker confidently predicts as a positive result.

Formally, as shown in Algorithm~\ref{alg:single_tool}, we set a threshold ${\tau}_s$ and fetch all $r_i$ with $\mathrm{score}(q,r_i)>{\tau}_s$. If such $r_i$ does not exist, we fetch $r_1$ instead. Then, we construct a set of tools $\mathcal{X}$ based on all the fetched $r_i$. Finally, we split $R$ into two lists $F_1$ and $F_2$ based on whether $\mathrm{tool}(r_i)$ is in $\mathcal{X}$ or not, and obtain $F$ by concatenating $F_1$ and $F_2$.

We also notice that some of the correct APIs may be located outside $T$, especially for unseen tools (see Section~\ref{sec:ablation} for details). Thus, we try to search potential correct APIs outside $T$ and build an extended API list $F_1$ for unseen tools. Specifically, for each unseen tool $x \in \mathcal{X}$, we search the entire tool library for the APIs which satisfy $\mathrm{tool}(\cdot)=x$ to build an extended $F_1$. Subsequently, we use the cross-encoder reranker to rerank $F_1$ again before concatenating it with $F_2$.

\begin{algorithm}[t]
\caption{Hierarchy-Aware Reranking for Multi-Tool Queries \label{alg:multi_tool}}
{\bf Input:}
user query $q$, reranked results $R=[r_1, \dots, r_l]$ \\
{\bf Output:}
final reranked results $F=[f_1, \dots, f_l]$
\begin{algorithmic}[1]
\State Construct a graph $G=(V,E)$ where $V=\{r_1,\dots,r_l\}$ and $E=\emptyset$
\For {$i\leftarrow 1$ to $l-1$}
    \For {$j\leftarrow i+1$ to $l$}
        \If {$\mathrm{tool}(r_i)=\mathrm{tool}(r_j)$ or $\mathrm{sim}(r_i,r_j) > {\tau}_m$}
            \State Add edge $\langle r_i,r_j\rangle$ to $E$
        \EndIf
    \EndFor
\EndFor
\State $\mathcal{S}\leftarrow\emptyset$
\For {each connected component $G'$ in $G$}
    \State Fetch at most $n$ results $r_{i_1},\dots,r_{i_{n^{\prime}}}$ with the best relevance scores $\mathrm{score}(q,\cdot)$ in $G'$
    \State $\mathcal{S}\leftarrow \mathcal{S}\cup\{r_{i_1},\dots,r_{i_{n^{\prime}}}\}$
\EndFor
\State $F1\leftarrow F2 \leftarrow [\ ]$
\For {$i\leftarrow 1$ to $l$}
    \If {$r_i\in\mathcal{S}$}
        \State $F1\leftarrow F1+[r_i]$
    \Else
        \State $F2\leftarrow F2+[r_i]$
    \EndIf
\EndFor
\State $F\leftarrow F1+F2$
\State \Return $F$
\end{algorithmic}
\end{algorithm}

\paragraph{Hierarchy-Aware Reranking for multi-tool queries.} For multi-tool queries, it is perferable for the fine-grained results to belong to functionally different tools. If a query can be resolved using multiple tools but the tools are functionally similar, we consider that it may probably be resolved using a single tool. Thus, we construct a graph to represent the hierarchical and semantic relations among the retrieval results and choose diverse results based on the graph.

Formally, as shown in Algorithm~\ref{alg:multi_tool}, we construct a graph $G=(V, E)$ to represent the hierarchical and semantic relations among the retrieval results, where $V$ includes all results in $R$. For each pair of results $\langle r_i, r_j\rangle$, we add an edge between them if they belong to the same tool or their semantic similarity $\mathrm{sim}(r_i, r_j)$ is greater than a threshold ${\tau}_m$. Then, we construct a set $\mathcal{S}$ which contains diverse retrieval results. For each connected component $G'$ in $G$, we fetch at most $n$ results $r_{i_1},\dots,r_{i_{n^{\prime}}}$ $(n^{\prime}\leq n)$ with the best relevance scores $\mathrm{score}(q,\cdot)$ and add them to $\mathcal{S}$. Finally, we split $R$ into $F_1$ and $F_2$ based on     whether $r_i$ is in $\mathcal{S}$ or not, and obtain $F$ by concatenating $F_1$ and $F_2$.

\begin{table}
\centering
\small
\begin{tabular}{cccr}
\toprule
\multicolumn{3}{c}{\textbf{Split}} & \textbf{\# of Queries} \\\midrule
\multicolumn{3}{c}{Train} & 169,287 \\
\multicolumn{3}{c}{Dev} & 600 \\\midrule
\multirow{6.5}{*}{Test} & \multirow{3}{*}{Seen} & I1-Inst & 100 \\
& & I2-Inst & 100 \\
& & I3-Inst & 100 \\\cmidrule(lr){2-4}
& \multirow{3}{*}{Unseen} & I1-Tool & 100 \\
& & I1-Cat & 100 \\
& & I2-Cat & 100 \\
\bottomrule
\end{tabular}
\caption{Statistics of the datasets used in our experiments. Following~\citet{qin2023toolllm}, we split the test dataset into 6 subsets. ``Seen'' and ``Unseen'' denote that the queries are related to tools seen or unseen in the training dataset, respectively.}
\label{tab:data_stats}
\end{table}

\section{Experiments}
\label{sec:experiments}

\subsection{Setup}
\label{sec:setup}

\paragraph{Data preparation.} We mainly conduct the experiments on the ToolBench~\citep{qin2023toolllm} dataset to validate the effectiveness of our proposed ToolRerank. Following~\citet{qin2023toolllm}, the test dataset is split into 6 subsets (I1-Inst, I2-Inst, I3-Inst, I1-Tool, I1-Cat and I2-Cat), each of which consists of 100 user queries.

To better evaluate the effectiveness of ToolRerank for unseen tools, we follow~\citet{qin2023toolllm} and remove some tools and categories from the original training data of ToolBench, making the related APIs for queries in I1-Tool, I1-Cat and I2-Cat unseen in the training data. Thus, we use \emph{unseen test datasets} to represent I1-Tool, I1-Cat and I2-Cat, and \emph{seen test datasets} to represent the remaining three test datasets.
Moreover, we also sample a development set with 600 user queries from the original training data.
The statistics of the datasets used in our experiments are presented in Table~\ref{tab:data_stats}.

\begin{table}
\centering
\small
\setlength{\tabcolsep}{2pt}
\begin{tabular}{cc}
\toprule
\textbf{Hyperparameter} & \textbf{Search Grid} \\\midrule
$m_s$ & \textbf{10}, 30, 50 \\
$m_u$ & 10, 30, \textbf{50} \\
${\tau}_s$ & 0.6, 0.65, 0.7, 0.75, 0.8, \textbf{0.85}, 0.9 \\
${\tau}_m$ & 0.6, 0.65, \textbf{0.7}, 0.75, 0.8, 0.85, 0.9 \\
$n$ & 2, \textbf{3}, 4 \\
\bottomrule
\end{tabular}
\caption{The hyperparameter search grid for Adaptive Truncation and Hierarchy-Aware Reranking. The best found hyperparameters are marked in \textbf{bold}.}
\label{tab:search_grid}
\end{table}

\paragraph{Baselines.} We compare ToolRerank with the following baselines:
\begin{enumerate}
\item BM25~\citep{robertson2009probabilistic}: The retrieval results are directly generated using a BM25 retriever.
\item DPR~\citep{qin2023toolllm}: The retrieval results are directly generated using a dual-encoder retriever. This is also the retrieval method proposed in~\citet{karpukhin2020dense}.
\item Rerank-$m$~\citep{nogueira2019passage}: The top-$m$ coarse-grained retrieval results are reranked by a cross-encoder reranker.\footnote{This reranker is trained on the sampled positive and hard negative pairs and is also used in our proposed ToolRerank.} The value of $m$ is set to \{10, 30, 50\} in our experiments.
\end{enumerate}

Note that we do not compare our method with the LLM-based retrievers~\citep{paranjape2023art} due to the inefficiency of using LLMs to retrieve suitable APIs from over 16,000 APIs provided in ToolBench~\citep{qin2023toolllm}.

\begin{table*}
\centering
\small
\setlength{\tabcolsep}{2pt}
\begin{tabular}{lcccccccccccccccccc}
\toprule
\multirow{4}{*}{\textbf{Method}} & \multicolumn{8}{c}{\textbf{Seen}} & \multicolumn{8}{c}{\textbf{Unseen}} & \multicolumn{2}{c}{\textbf{All}} \\\cmidrule(lr){2-9}\cmidrule(lr){10-17}\cmidrule(lr){18-19}
& \multicolumn{2}{c}{I1-Inst} & \multicolumn{2}{c}{I2-Inst} & \multicolumn{2}{c}{I3-Inst} & \multicolumn{2}{c}{Average} & \multicolumn{2}{c}{I1-Tool} & \multicolumn{2}{c}{I1-Cat} & \multicolumn{2}{c}{I2-Cat} & \multicolumn{2}{c}{Average} & \multicolumn{2}{c}{Average} \\\cmidrule(lr){2-3}\cmidrule(lr){4-5}\cmidrule(lr){6-7}\cmidrule(lr){8-9}\cmidrule(lr){10-11}\cmidrule(lr){12-13}\cmidrule(lr){14-15}\cmidrule(lr){16-17}\cmidrule(lr){18-19}
& N & R & N & R & N & R & N & R & N & R & N & R & N & R & N & R & N & R \\\midrule
BM25 & 45.2 & 48.2 & 40.9 & 44.3 & 33.6 & 35.7 & 39.9 & 42.7 & 59.3 & 63.9 & 53.6 & 55.7 & 33.8 & 35.5 & 48.9 & 51.7 & 44.4 & 47.2 \\
DPR & 87.5 & 90.8 & 81.7 & 85.4 & 80.7 & 82.7 & 83.3 & 86.3 & 74.5 & 80.5 & 59.2 & 64.1 & 42.4 & 47.6 & 58.7 & 64.0 & 71.0 & 75.2 \\\midrule
Rerank-10 & 92.1 & 94.2 & 84.2 & 87.3 & 84.9 & 85.1 & 87.0 & 88.9 & 80.3 & 82.9 & 69.9 & 68.7 & 56.5 & 57.3 & 68.9 & 69.6 & 78.0 & 79.2 \\
Rerank-30 & 89.1 & 91.7 & 80.6 & 81.8 & 84.2 & 84.0 & 84.6 & 85.8 & 82.5 & 85.6 & 75.6 & 76.1 & 61.1 & 62.4 & 73.1 & 74.7 & 78.8 & 80.3 \\
Rerank-50 & 88.4 & 91.4 & 79.6 & 80.2 & 82.6 & 81.6 & 83.5 & 84.4 & 80.6 & 82.6 & 75.5 & 77.1 & 62.5 & 63.4 & 72.8 & 74.4 & 78.2 & 79.4 \\\midrule
ToolRerank & \textbf{92.2} & \textbf{95.0} & \textbf{84.4} & \textbf{88.2} & \textbf{85.5} & \textbf{85.6} & \textbf{87.3} & \textbf{89.6} & \textbf{85.3} & \textbf{88.2} & \textbf{79.0} & \textbf{81.6} & \textbf{66.1} & \textbf{66.4} & \textbf{76.8} & \textbf{78.7} & \textbf{82.1} & \textbf{84.2} \\
\bottomrule
\end{tabular}
\caption{Comparison of the quality of the retrieval results on the six test datasets of the ToolBench dataset. ``N'' and ``R'' denote ``NDCG@5'' and ``Recall@5'', respectively.}
\label{tab:main_result_1}
\end{table*}

\paragraph{Implementation details.} The retriever, the reranker and the classifier are initialized with \texttt{bert-base-uncased}~\citep{devlin2019bert}. The retriever and the classifier are directly trained on the training dataset. For the reranker, we sample positive and hard negative pairs to construct its training dataset. Specifically, the positive pairs are obtained from the golden annotations in the training dataset, while the hard negative pairs are sampled from the coarse-grained retrieval results generated by the retriever.

We search the hyperparameters used in Adaptive Truncation and Hierarchy-Aware Reranking ($m_s$, $m_u$, ${\tau}_s$, ${\tau}_m$ and $n$) by grid search on the development set. The hyperparameter search grid and the found best hyperparameters are shown in Table~\ref{tab:search_grid}. 
For the fine-grained retrieval results, we follow~\citet{qin2023toolllm} and give the top $k=5$ results in the final reranked results $F$ to the LLM.

\subsection{Main Results}
\label{sec:main_results}

First, we compare the retrieval performance between the baselines and our proposed ToolRerank on the six test datasets of the ToolBench dataset. We use NDCG@5~\citep{jrvelin2002cumulated} and Recall@5 as the evaluation metrics to evaluate the retrieval performance. As presented in Table~\ref{tab:main_result_1}, ToolRerank consistently outperforms the baselines on all six test datasets.

On \emph{seen test datasets}, the baseline Rerank-10 outperforms DPR, but Rerank-30 and Rerank-50 underperform DPR, which indicates that giving more candidates to the reranker may have a negative impact on the retrieval performance. This is because the majority (91.9\%) of the correct APIs in these datasets appear in the top-10 coarse-grained results, and thus adding more candidates will inject more noise into the retrieval results.

On \emph{unseen test datasets}, giving more candidates to the reranker improves the retrieval performance. Specifically, among Rerank-\{10, 30, 50\}, Rerank-30 performs best on I1-Tool, and Rerank-50 performs best on I1-Cat and I2-Cat. This is because a smaller percentage (70.2\%) of correct APIs appear in the top-10 coarse-grained results for \emph{unseen test datasets}. Thus, adding more candidates may increase the possibility that the reranker can find the correct APIs.

\begin{table}
\centering
\small
\begin{tabular}{lcc}
\toprule
\textbf{Method} & \textbf{Pass Rate} & \textbf{Win Rate} \\\midrule
Oracle & 61.0 & 68.3 \\\midrule
BM25 & 58.8 & 50.8 \\
DPR & 57.2 & 49.8 \\\midrule
Rerank-10 & 60.2 & 53.7 \\
Rerank-30 & 59.8 & 53.7 \\
Rerank-50 & 59.8 & 55.2 \\\midrule
ToolRerank & \textbf{61.5} & \textbf{57.0} \\
\bottomrule
\end{tabular}
\caption{Effect of reranking on the execution results generated by the LLM.}
\label{tab:execute}
\end{table}

With Adaptive Truncation and Hierarchy-Aware Reranking, our proposed ToolRerank consistently makes further improvement over the baselines. Specifically, with Adaptive Truncation, ToolRerank can choose the number of candidates adaptively based on whether the query is related to seen or unseen tools. Generally, when the query is related to unseen tools, fewer candidates are given to the reranker. Otherwise, more candidates are given to the reranker. This leads to decent performance for both seen and unseen tools. With Hierarchy-Aware Reranking, ToolRerank can make the retrieval results more concentrated for single-tool queries and more diverse for multi-tool queries, which further improves the retrieval performance. As a result, ToolRerank outperforms Rerank-\{10, 30, 50\} by 5.0, 3.9 and 4.8 Recall@5 points on average on all six test datasets, respectively.

\begin{table}
\centering
\small
\begin{tabular}{cccccc}
\toprule
& $\bm{m_s}$ & $\bm{m_u}$ & \textbf{Seen} & \textbf{Unseen} & \textbf{All} \\\midrule
\multirow{3}{*}{\makecell[c]{$m_s<m_u$}} & 10 & 50 & \textbf{89.6} & \textbf{78.7} & \textbf{84.2} \\
& 10 & 30 & 89.6 & 77.8 & 83.7 \\
& 30 & 50 & 87.3 & 77.5 & 82.4 \\\midrule
\multirow{3}{*}{$m_s=m_u$} & 10 & 10 & 89.6 & 74.5 & 82.0 \\
& 30 & 30 & 87.2 & 76.7 & 81.9 \\
& 50 & 50 & 85.6 & 76.3 & 81.0 \\\midrule
\multirow{3}{*}{$m_s>m_u$} & 50 & 10 & 85.7 & 71.3 & 78.5 \\
& 30 & 10 & 87.1 & 72.2 & 79.7 \\
& 50 & 30 & 85.7 & 75.4 & 80.6 \\
\bottomrule
\end{tabular}
\caption{Effect of Adaptive Truncation on Recall@5.}
\label{tab:ablation}
\end{table}

\begin{table*}
\centering
\small
\begin{tabular}{lccccccccc}
\toprule
\multirow{2.5}{*}{\textbf{Variant}} & \multicolumn{4}{c}{\textbf{Single-Tool}} & \multicolumn{4}{c}{\textbf{Multi-Tool}} & \textbf{All} \\\cmidrule(lr){2-5}\cmidrule(lr){6-9}\cmidrule(lr){10-10}
& I1-Inst & I1-Tool & I1-Cat & Average & I2-Inst & I3-Inst & I2-Cat & Average & Average \\\midrule
ToolRerank & \textbf{95.0} & 88.2 & 81.6 & 88.3 & \textbf{88.2} & \textbf{85.6} & \textbf{66.4} & \textbf{80.1} & \textbf{84.2} \\\midrule
ToolRerank$_\text{none}$ & 94.2 & 87.5 & 77.8 & 86.5 & 86.9 & 84.8 & 65.7 & 79.1 & 82.8 \\
ToolRerank$_\text{single}$ & \textbf{95.0} & \textbf{89.7} & \textbf{81.9} & \textbf{88.9} & 85.4 & 83.9 & 64.8 & 78.0 & 83.5 \\
ToolRerank$_\text{multi}$ & 93.4 & 85.6 & 76.6 & 85.2 & \textbf{88.2} & \textbf{85.6} & \textbf{66.4} & \textbf{80.1} & 82.6 \\\midrule
ToolRerank$_\text{oracle}$ & 95.0 & 89.7 & 81.9 & 88.9 & 88.2 & 85.6 & 66.4 & 80.1 & 84.5 \\
\bottomrule
\end{tabular}
\caption{Effect of Hierarchy-Aware Reranking on Recall@5.}
\label{tab:ablation_classifier}
\end{table*}

\begin{table}
\centering
\small
\setlength{\tabcolsep}{3pt}
\begin{tabular}{cccccc}
\toprule
\multicolumn{2}{c}{\textbf{Extended}} & \multirow{2.5}{*}{\textbf{I1-Inst}} & \multirow{2.5}{*}{\textbf{I1-Tool}} & \multirow{2.5}{*}{\textbf{I1-Cat}} & \multirow{2.5}{*}{\textbf{Average}} \\\cmidrule(lr){1-2}
Seen & Unseen & & & & \\\midrule
No & Yes & \textbf{95.0} & \textbf{88.2} & \textbf{81.6} & \textbf{88.3} \\\midrule
No & No & \textbf{95.0} & 86.9 & 81.1 & 87.7 \\
Yes & Yes & 94.5 & 87.6 & 80.9 & 87.7 \\
Yes & No & 94.5 & 86.3 & 80.4 & 87.1 \\
\bottomrule
\end{tabular}
\caption{Effect of the extended API list on Recall@5 for single-tool queries.}
\label{tab:ablation_extend}
\end{table}

\subsection{Effect of Reranking on Execution Results generated by LLM}

In this section, we examine how reranking affects the execution results generated by the LLM. Following~\citet{qin2023toolllm}, we adopt the ToolLLaMA-7b\footnote{\url{https://huggingface.co/ToolBench/ToolLLaMA-7b-v1}} model as the backbone LLM and generate the execution results using the DFSDT algorithm. We use the Pass Rate and the Win Rate metrics~\citep{qin2023toolllm} to evaluate the quality of the execution results generated by the LLM. Specifically, the Pass Rate is the proportion of user queries for which the LLM can provide a valid solution, and the Win Rate is the proportion of user queries for which the LLM can provide a solution better than the baseline solution (which is produced using ChatGPT~\citep{openaichatgptblog} and ReAct~\citep{yao2023react}).

The experimental results are shown in Table~\ref{tab:execute}. First, we find that using a better retriever does not necessarily lead to better execution results, as demonstrated by the inferior performance of DPR compared with BM25 on both metrics. Second, Rerank-\{10, 30, 50\} and ToolRerank outperform BM25 and DPR on both metrics, suggesting that reranking improves the quality of the execution results. Finally, compared with Rerank-\{10, 30, 50\}, ToolRerank can further improve the quality of the execution results, thereby highlighting the advantage of ToolRerank for tool retrieval.

However, even with ToolRerank, the Win Rate remains significantly lower than that when the oracle APIs are provided. This suggests that there is still room for improvement of the reranking method.

\subsection{Ablation Studies}
\label{sec:ablation}

In this section, we conduct further ablation studies to assess the efficacy of different components of ToolRerank. For simplicity, we only report Recall@5 in this section, as the order of the APIs given to the LLM does not significantly affect the quality of the execution results. In fact, we conduct an additional experiment where the APIs given to the LLM are randomly shuffled, resulting in a 16.5 point decrease in NDCG@5, but no significant impact on Pass Rate and Win Rate.

\paragraph{Effect of Adaptive Truncation.} As demonstrated in Table~\ref{tab:ablation}, to investigate the effect of Adaptive Truncation, we conduct further experiments using different truncation strategies. On the one hand, we find that the performance deteriorates when a fixed number of candidates are given to the reranker ($m_s=m_u$), showing the effectiveness of Adaptive Truncation for tool retrieval. On the other hand, if we set $m_s>m_u$ for Adaptive Truncation, the performance will become even worse than that when $m_s=m_u$. Therefore, we conclude that Adaptive Truncation is effective only when $m_s<m_u$.

\paragraph{Effect of Hierarchy-Aware Reranking.} As shown in Table~\ref{tab:ablation_classifier}, to explore the effectivess of Hierarchy-Aware Reranking, we conduct additional experiments using different reranking algorithms.

First, the performance on both single- and multi-tool test datasets decreases if we do not use Hierarchy-Aware Reranking (ToolRerank$_\text{none}$), highlighting the effectiveness of Hierarchy-Aware Reranking for tool retrieval.

Second, when applying only one of the reranking algorithms to all test datasets (ToolRerank$_{\text{single}}$ and ToolRerank$_\text{multi}$), the performance decreases when the used algorithm does not match the query type (single-tool or multi-tool).
This suggests that the Hierarchy-Aware Reranking is effective only when the used algorithm matches the query type, highlighting the importance of using a classifier to distinguish between single- and multi-tool queries.

Finally, to explore the upper bound of ToolRerank, we also try to use the oracle classification result to choose the appropriate reranking algorithm (ToolRerank$_\text{oracle}$). The result shows that the retrieval performance only increases by 0.3 points on average, which further demonstrates the effectiveness of ToolRerank even when some queries are misclassified. In fact, the average accuracy of the classifier used in our experiments is 93.3\%.

\begin{figure}[t]
\begin{center}
\includegraphics[scale=0.7]{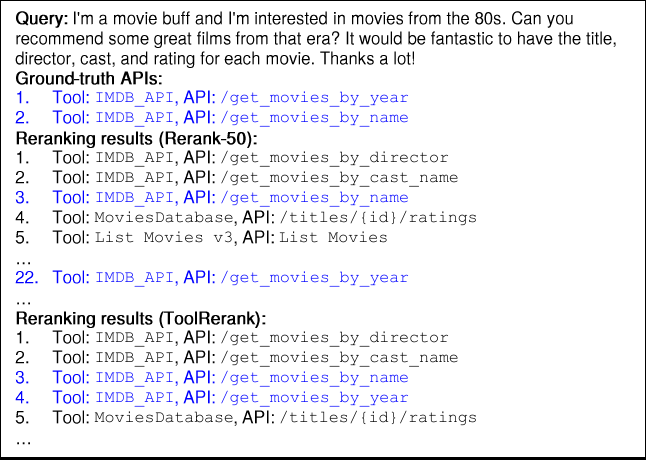} 
\caption{Case study of a single-tool query in I1-Cat. Correct APIs are highlighted in \textcolor{blue}{blue}.}
\label{fig:casestudy1}
\end{center}
\end{figure}

\paragraph{Effect of the extended API list.} As presented in Table~\ref{tab:ablation_extend}, we also conduct experiments to investigate the effect of the extended API list $F_1$ for single-tool queries. First, we find that the absence of the extended $F_1$ decreases the performance on single-tool \emph{unseen test datasets} (I1-Tool and I1-Cat). This suggests that we may find potential correct APIs outside $T$ using the extended $F_1$. Second, the performance on all single-tool test datasets decreases if we build the extended $F_1$ for seen tools. This may be because only a small percentage (2.6\%) of the correct APIs of the seen tools appear outside $T$ and thus building the extended $F_1$ for seen tools may inject more noise into the retrieval results. In contrast, a higher percentage (7.7\%) of correct APIs of the unseen tools appear outside $T$.

We also find that building the extended $F_1$ for multi-tool queries harms the retrieval performance, since we are unable to find any correct API outside $T$ which belongs to the same tool as any API listed in $F_1$ for the multi-tool test datasets.

\subsection{Case Studies}

In this section, we also conduct two case studies to further show how ToolRerank improves the retrieval performance.

Figure~\ref{fig:casestudy1} presents the retrieval results for a single-tool query in I1-Cat. Rerank-50 ranks the correct API \texttt{/get\_movies\_by\_name} (which belongs to \texttt{IMDB\_API}) at the 22nd position. However, ToolRerank suggests that the correct APIs should belong to \texttt{IMDB\_API} and lists \texttt{/get\_movies\_by\_name} in $F_1$. Thus, \texttt{/get\_movies\_by\_name} is ranked at the 4th position, which can be given to the LLM when $k$ is set to 5.

\begin{figure}[t]
\begin{center}
\includegraphics[scale=0.7]{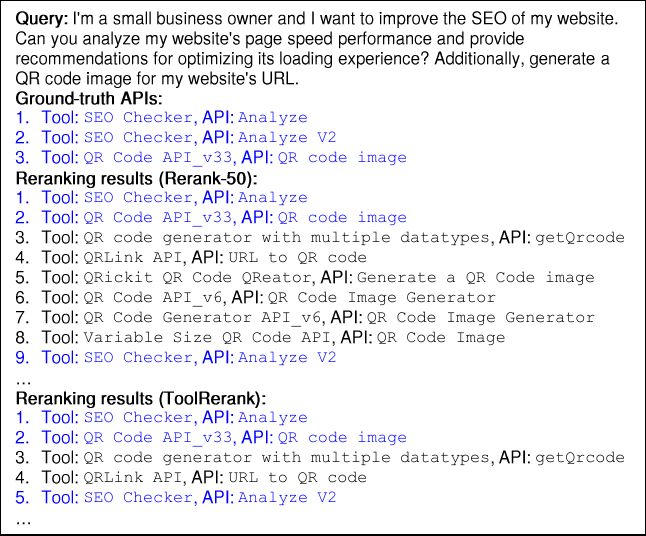} 
\caption{Case study of a multi-tool query in I2-Cat. Correct APIs are highlighted in \textcolor{blue}{blue}.}
\label{fig:casestudy2}
\end{center}
\end{figure}

Figure~\ref{fig:casestudy2} presents the retrieval results for a multi-tool query in I2-Cat. Rerank-50 ranks 7 different APIs related to QR code from the 2nd to the 8th position. Consequently, the correct API \texttt{Analyze V2} (which belongs to \texttt{SEO Checker}) is ranked at the 9th position. However, since these APIs related to QR code are functionally similar to each other, ToolRerank only lists 3 of them in $F_1$. As a result, \texttt{Analyze V2} is ranked at the 5th position and can be given to the LLM.

\section{Related Work}

This work is highly related to the following two lines of research: (1) tool learning and (2) reranking for information retrieval.

\subsection{Tool Learning}

Tool learning aims to extend the capabilities of large language models (LLMs) with external tools~\citep{schick2023toolformer,chen2023chatcot,yao2023react}. For example, an LLM augmented with tools may be able to leverage the latest information~\citep{yang2023chatgpt,liu2023webglm}, perform complex arithmetic calculations~\citep{cobbe2021training,gao2023pal} or process multi-modal information~\citep{shen2023hugginggpt,lu2023chameleon}.

However, it is challenging to support a large number of tools since we need to provide the API documents in the input context~\citep{qin2023tool}. To address this challenge, \citet{hao2023toolkengpt} use the examples of tool usage to fine-tune the extra token embeddings which represent the actions of calling the APIs.

Besides, various studies have proposed using retrieval-based methods to support a large number of tools~\citep{paranjape2023art,patil2023gorilla,qin2023toolllm}. Retrieval-based methods can be combined with advanced fine-tuning methods to reach even better performance~\citep{gao2023confucius}.
However, these studies do not incorporate reranking methods to refine the retrieval results or consider the unique features of tool retrieval.
In contrast, we propose a specialized reranking method for tool retrieval in this work, which handles seen and unseen tools separately and takes the hierarchy of the tool library into account.

\subsection{Reranking for Information Retrieval}

In the research field of information retrieval, reranking aims to refine the coarse-grained retrieval results to enhance the retrieval performance using more computationally intensive models~\citep{guo2016deep,xiong2017endtoend,nogueira2019passage,yan2019idst}. A common approach to reranking is using a cross-encoder reranker to rerank a fixed number of candidates generated by the retriever~\citep{nogueira2019passage}.

Furthermore, previous studies have proposed more sophisticated reranking methods to further improve the retrieval performance.
For example, \citet{ren2021rocketqav2} and~\citet{zhang2022adversarial} jointly train the retriever and the reranker to improve both models. \citet{zhou2023towards} generate more diverse hard negative pairs and utilize the label noise to enhance the robustness of the reranker. \citet{wang2022text} pretrain the model on a large-scale text pair dataset using a weakly supervised contrastive learning objective before fine-tuning the model for reranking.

Inspired by the reranking methods for information retrieval, we propose a specialized reranking method for tool retrieval in this work, which is also proven effective through experimentation. In the future, we may benefit from information retrieval methods to further improve our method.

\section{Conclusion}

In this work, we propose ToolRerank, an adaptive and hierarchy-aware reranking method for tool retrieval, which handles seen and unseen tools separately and takes the hierarchy of the tool library into account. Experimental results show that ToolRerank can improve the quality of the retrieval results.
Additionally, the APIs provided by ToolRerank enable the LLM to generate better execution results.

\section{Ethics Statement}

We consider that there may be some potential risk if our proposed ToolRerank is misused. For example, if the tool library contains harmful tools, ToolRerank may make the harmful tools easier to be retrieved when a harmful user query is given. In this case, the harmful tools may become easier to be used by malicious users.

\section{Acknowledgments}

This work is supported by the National Key R\&D Program of China (2022ZD0160502), the National Natural Science Foundation of China (No. 61925601, 62276152) and Xiaomi AI Lab. We thank all anonymous reviewers for their valuable comments and suggestions on this work.

\nocite{*}
\section{Bibliographical References}\label{sec:reference}

\bibliographystyle{lrec-coling2024-natbib}
\bibliography{lrec-coling2024-example}

\begin{thebibliography}{33}
\expandafter\ifx\csname natexlab\endcsname\relax\def\natexlab#1{#1}\fi

\bibitem[{Chen et~al.(2023)Chen, Zhou, Zhang, Gong, Zhao, and
  Wen}]{chen2023chatcot}
Zhipeng Chen, Kun Zhou, Beichen Zhang, Zheng Gong, Xin Zhao, and Ji{-}Rong Wen.
  2023.
\newblock \href {https://aclanthology.org/2023.findings-emnlp.985} {Chatcot:
  Tool-augmented chain-of-thought reasoning on chat-based large language
  models}.
\newblock In \emph{Findings of EMNLP}, pages 14777--14790.

\bibitem[{Cobbe et~al.(2021)Cobbe, Kosaraju, Bavarian, Chen, Jun, Kaiser,
  Plappert, Tworek, Hilton, Nakano, Hesse, and Schulman}]{cobbe2021training}
Karl Cobbe, Vineet Kosaraju, Mohammad Bavarian, Mark Chen, Heewoo Jun, Lukasz
  Kaiser, Matthias Plappert, Jerry Tworek, Jacob Hilton, Reiichiro Nakano,
  Christopher Hesse, and John Schulman. 2021.
\newblock \href {https://arxiv.org/abs/2110.14168} {Training verifiers to solve
  math word problems}.
\newblock \emph{arXiv preprint arXiv:2110.14168}.

\bibitem[{Devlin et~al.(2019)Devlin, Chang, Lee, and
  Toutanova}]{devlin2019bert}
Jacob Devlin, Ming{-}Wei Chang, Kenton Lee, and Kristina Toutanova. 2019.
\newblock \href {https://aclanthology.org/N19-1423} {{BERT:} pre-training of
  deep bidirectional transformers for language understanding}.
\newblock In \emph{Proc. of NAACL-HLT}, pages 4171--4186.

\bibitem[{Gao et~al.(2023{\natexlab{a}})Gao, Madaan, Zhou, Alon, Liu, Yang,
  Callan, and Neubig}]{gao2023pal}
Luyu Gao, Aman Madaan, Shuyan Zhou, Uri Alon, Pengfei Liu, Yiming Yang, Jamie
  Callan, and Graham Neubig. 2023{\natexlab{a}}.
\newblock \href {https://proceedings.mlr.press/v202/gao23f.html} {{PAL}:
  Program-aided language models}.
\newblock In \emph{Proc. of ICML}, volume 202, pages 10764--10799.

\bibitem[{Gao et~al.(2023{\natexlab{b}})Gao, Shi, Zhu, Fang, Xin, Ren, Chen,
  and Ma}]{gao2023confucius}
Shen Gao, Zhengliang Shi, Minghang Zhu, Bowen Fang, Xin Xin, Pengjie Ren,
  Zhumin Chen, and Jun Ma. 2023{\natexlab{b}}.
\newblock \href {https://arxiv.org/abs/2308.14034} {Confucius: Iterative tool
  learning from introspection feedback by easy-to-difficult curriculum}.
\newblock \emph{arXiv preprint arXiv:2308.14034}.

\bibitem[{Guo et~al.(2016)Guo, Fan, Ai, and Croft}]{guo2016deep}
J.~Guo, Yixing Fan, Qingyao Ai, and W.~Bruce Croft. 2016.
\newblock \href {https://dl.acm.org/doi/10.1145/2983323.2983769} {A deep
  relevance matching model for ad-hoc retrieval}.
\newblock In \emph{Proc. of CIKM}, pages 55--64.

\bibitem[{Hao et~al.(2023)Hao, Liu, Wang, and Hu}]{hao2023toolkengpt}
Shibo Hao, Tianyang Liu, Zhen Wang, and Zhiting Hu. 2023.
\newblock \href {https://arxiv.org/abs/2305.11554} {Toolkengpt: Augmenting
  frozen language models with massive tools via tool embeddings}.
\newblock \emph{arXiv preprint arXiv:2305.11554}.

\bibitem[{Hsieh et~al.(2023)Hsieh, Chen, Li, Fujii, Ratner, Lee, Krishna, and
  Pfister}]{hsieh2023tool}
Cheng-Yu Hsieh, Sibei Chen, Chun-Liang Li, Yasuhisa Fujii, Alexander~J. Ratner,
  Chen-Yu Lee, Ranjay Krishna, and Tomas Pfister. 2023.
\newblock \href {https://arxiv.org/abs/2308.00675} {Tool documentation enables
  zero-shot tool-usage with large language models}.
\newblock \emph{arXiv preprint arXiv:2308.00675}.

\bibitem[{Humeau et~al.(2020)Humeau, Shuster, Lachaux, and
  Weston}]{humeau2020poly}
Samuel Humeau, Kurt Shuster, Marie{-}Anne Lachaux, and Jason Weston. 2020.
\newblock \href {https://openreview.net/forum?id=SkxgnnNFvH} {Poly-encoders:
  Architectures and pre-training strategies for fast and accurate
  multi-sentence scoring}.
\newblock In \emph{Proc. of ICLR}.

\bibitem[{J{\"a}rvelin and Kek{\"a}l{\"a}inen(2002)}]{jrvelin2002cumulated}
Kalervo J{\"a}rvelin and Jaana Kek{\"a}l{\"a}inen. 2002.
\newblock \href {https://dl.acm.org/doi/10.1145/582415.582418} {Cumulated
  gain-based evaluation of ir techniques}.
\newblock \emph{ACM Trans. Inf. Syst.}, 20:422--446.

\bibitem[{Karpukhin et~al.(2020)Karpukhin, Oguz, Min, Lewis, Wu, Edunov, Chen,
  and Yih}]{karpukhin2020dense}
Vladimir Karpukhin, Barlas Oguz, Sewon Min, Patrick S.~H. Lewis, Ledell Wu,
  Sergey Edunov, Danqi Chen, and Wen{-}tau Yih. 2020.
\newblock \href {https://aclanthology.org/2020.emnlp-main.550} {Dense passage
  retrieval for open-domain question answering}.
\newblock In \emph{Proc. of EMNLP}, pages 6769--6781.

\bibitem[{Lin et~al.(2022)Lin, Mihaylov, Artetxe, Wang, Chen, Simig, Ott,
  Goyal, Bhosale, Du, Pasunuru, Shleifer, Koura, Chaudhary, O'Horo, Wang,
  Zettlemoyer, Kozareva, Diab, Stoyanov, and Li}]{lin2022fewshot}
Xi~Victoria Lin, Todor Mihaylov, Mikel Artetxe, Tianlu Wang, Shuohui Chen,
  Daniel Simig, Myle Ott, Naman Goyal, Shruti Bhosale, Jingfei Du, Ramakanth
  Pasunuru, Sam Shleifer, Punit~Singh Koura, Vishrav Chaudhary, Brian O'Horo,
  Jeff Wang, Luke Zettlemoyer, Zornitsa Kozareva, Mona~T. Diab, Veselin
  Stoyanov, and Xian Li. 2022.
\newblock \href {https://aclanthology.org/2022.emnlp-main.616} {Few-shot
  learning with multilingual generative language models}.
\newblock In \emph{Proc. of {EMNLP}}, pages 9019--9052.

\bibitem[{Liu et~al.(2023)Liu, Lai, Yu, Xu, Zeng, Du, Zhang, Dong, and
  Tang}]{liu2023webglm}
Xiao Liu, Hanyu Lai, Hao Yu, Yifan Xu, Aohan Zeng, Zhengxiao Du, Peng Zhang,
  Yuxiao Dong, and Jie Tang. 2023.
\newblock \href {https://dl.acm.org/doi/10.1145/3580305.3599931} {{WebGLM}:
  Towards an efficient web-enhanced question answering system with human
  preferences}.
\newblock In \emph{Proc. of SIGKDD}, pages 4549--4560.

\bibitem[{Lu et~al.(2023)Lu, Peng, Cheng, Galley, Chang, Wu, Zhu, and
  Gao}]{lu2023chameleon}
Pan Lu, Baolin Peng, Hao Cheng, Michel Galley, Kai-Wei Chang, Ying~Nian Wu,
  Song-Chun Zhu, and Jianfeng Gao. 2023.
\newblock \href {https://arxiv.org/abs/2304.09842} {Chameleon: Plug-and-play
  compositional reasoning with large language models}.
\newblock \emph{arXiv preprint arXiv:2304.09842}.

\bibitem[{Nogueira and Cho(2019)}]{nogueira2019passage}
Rodrigo~Frassetto Nogueira and Kyunghyun Cho. 2019.
\newblock \href {http://arxiv.org/abs/1901.04085} {Passage re-ranking with
  {BERT}}.
\newblock \emph{arXiv preprint arXiv:1901.04085}.

\bibitem[{OpenAI(2022)}]{openaichatgptblog}
OpenAI. 2022.
\newblock \href {https://openai.com/blog/chatgpt} {Open{AI}: Introducing
  {ChatGPT}}.

\bibitem[{OpenAI(2023)}]{openai2023gpt4}
OpenAI. 2023.
\newblock \href {http://arxiv.org/abs/2303.08774} {Gpt-4 technical report}.

\bibitem[{Paranjape et~al.(2023)Paranjape, Lundberg, Singh, Hajishirzi,
  Zettlemoyer, and Ribeiro}]{paranjape2023art}
Bhargavi Paranjape, Scott~M. Lundberg, Sameer Singh, Hanna Hajishirzi, Luke
  Zettlemoyer, and Marco~Tulio Ribeiro. 2023.
\newblock \href {https://arxiv.org/abs/2303.09014} {Art: Automatic multi-step
  reasoning and tool-use for large language models}.
\newblock \emph{arXiv preprint arXiv:2303.09014}.

\bibitem[{Patel et~al.(2021)Patel, Bhattamishra, and Goyal}]{patel2021nlp}
Arkil Patel, Satwik Bhattamishra, and Navin Goyal. 2021.
\newblock \href {https://aclanthology.org/2021.naacl-main.168} {Are {NLP}
  models really able to solve simple math word problems?}
\newblock In \emph{Proc. of {NAACL-HLT}}, pages 2080--2094.

\bibitem[{Patil et~al.(2023)Patil, Zhang, Wang, and
  Gonzalez}]{patil2023gorilla}
Shishir~G. Patil, Tianjun Zhang, Xin Wang, and Joseph~E. Gonzalez. 2023.
\newblock \href {https://arxiv.org/abs/2305.15334} {Gorilla: Large language
  model connected with massive apis}.
\newblock \emph{arXiv preprint arXiv:2305.15334}.

\bibitem[{Qin et~al.(2023{\natexlab{a}})Qin, Hu, Lin, Chen, Ding, Cui, Zeng,
  Huang, Xiao, Han, Fung, Su, Wang, Qian, Tian, Zhu, Liang, Shen, Xu, Zhang,
  Ye, Li, Tang, Yi, Zhu, Dai, Yan, Cong, Lu, Zhao, Huang, Yan, Han, Sun, Li,
  Phang, Yang, Wu, Ji, Liu, and Sun}]{qin2023tool}
Yujia Qin, Shengding Hu, Yankai Lin, Weize Chen, Ning Ding, Ganqu Cui, Zheni
  Zeng, Yufei Huang, Chaojun Xiao, Chi Han, Yi~Ren Fung, Yusheng Su, Huadong
  Wang, Cheng Qian, Runchu Tian, Kunlun Zhu, Shi Liang, Xingyu Shen, Bokai Xu,
  Zhen Zhang, Yining Ye, Bo~Li, Ziwei Tang, Jing Yi, Yu~Zhu, Zhenning Dai, Lan
  Yan, Xin Cong, Ya-Ting Lu, Weilin Zhao, Yuxiang Huang, Jun-Han Yan, Xu~Han,
  Xian Sun, Dahai Li, Jason Phang, Cheng Yang, Tongshuang Wu, Heng Ji, Zhiyuan
  Liu, and Maosong Sun. 2023{\natexlab{a}}.
\newblock \href {https://arxiv.org/abs/2304.08354} {Tool learning with
  foundation models}.
\newblock \emph{arXiv preprint arXiv:2304.08354}.

\bibitem[{Qin et~al.(2023{\natexlab{b}})Qin, Liang, Ye, Zhu, Yan, Lu, Lin,
  Cong, Tang, Qian, Zhao, Tian, Xie, Zhou, Gerstein, Li, Liu, and
  Sun}]{qin2023toolllm}
Yujia Qin, Shihao Liang, Yining Ye, Kunlun Zhu, Lan Yan, Yaxi Lu, Yankai Lin,
  Xin Cong, Xiangru Tang, Bill Qian, Sihan Zhao, Runchu Tian, Ruobing Xie, Jie
  Zhou, Mark Gerstein, Dahai Li, Zhiyuan Liu, and Maosong Sun.
  2023{\natexlab{b}}.
\newblock \href {https://arxiv.org/abs/2307.16789v1} {{ToolLLM}: Facilitating
  large language models to master 16000+ real-world {APIs}}.
\newblock \emph{arXiv preprint arXiv:2307.16789v1}.

\bibitem[{Ren et~al.(2021)Ren, Qu, Liu, Zhao, She, Wu, Wang, and
  Wen}]{ren2021rocketqav2}
Ruiyang Ren, Yingqi Qu, Jing Liu, Wayne~Xin Zhao, Qiaoqiao She, Hua Wu, Haifeng
  Wang, and Ji{-}Rong Wen. 2021.
\newblock \href {https://aclanthology.org/2021.emnlp-main.224} {{RocketQAv2}:
  {A} joint training method for dense passage retrieval and passage
  re-ranking}.
\newblock In \emph{Proc. of EMNLP}, pages 2825--2835.

\bibitem[{Robertson and Zaragoza(2009)}]{robertson2009probabilistic}
Stephen~E. Robertson and Hugo Zaragoza. 2009.
\newblock \href {https://www.nowpublishers.com/article/Details/INR-019} {The
  probabilistic relevance framework: {BM25} and beyond}.
\newblock \emph{Found. Trends Inf. Retr.}, 3:333--389.

\bibitem[{Schick et~al.(2023)Schick, Dwivedi{-}Yu, Dess{\`{\i}}, Raileanu,
  Lomeli, Zettlemoyer, Cancedda, and Scialom}]{schick2023toolformer}
Timo Schick, Jane Dwivedi{-}Yu, Roberto Dess{\`{\i}}, Roberta Raileanu, Maria
  Lomeli, Luke Zettlemoyer, Nicola Cancedda, and Thomas Scialom. 2023.
\newblock \href {https://arxiv.org/abs/2302.04761} {Toolformer: Language models
  can teach themselves to use tools}.
\newblock \emph{arXiv preprint arXiv:2302.04761}.

\bibitem[{Shen et~al.(2023)Shen, Song, Tan, Li, Lu, and
  Zhuang}]{shen2023hugginggpt}
Yongliang Shen, Kaitao Song, Xu~Tan, Dongsheng Li, Weiming Lu, and Yueting
  Zhuang. 2023.
\newblock \href {https://arxiv.org/abs/2303.17580} {{HuggingGPT}: Solving {AI}
  tasks with chatgpt and its friends in huggingface}.
\newblock \emph{arXiv preprint arXiv:2303.17580}.

\bibitem[{Wang et~al.(2022)Wang, Yang, Huang, Jiao, Yang, Jiang, Majumder, and
  Wei}]{wang2022text}
Liang Wang, Nan Yang, Xiaolong Huang, Binxing Jiao, Linjun Yang, Daxin Jiang,
  Rangan Majumder, and Furu Wei. 2022.
\newblock \href {https://arxiv.org/abs/2212.03533} {Text embeddings by
  weakly-supervised contrastive pre-training}.
\newblock \emph{arXiv preprint arXiv:2212.03533}.

\bibitem[{Xiong et~al.(2017)Xiong, Dai, Callan, Liu, and
  Power}]{xiong2017endtoend}
Chenyan Xiong, Zhuyun Dai, Jamie Callan, Zhiyuan Liu, and Russell Power. 2017.
\newblock \href {https://dl.acm.org/doi/10.1145/3077136.3080809} {End-to-end
  neural ad-hoc ranking with kernel pooling}.
\newblock In \emph{Proc. of SIGIR}, pages 55--64.

\bibitem[{Yan et~al.(2019)Yan, Li, Wu, Xia, and Wang}]{yan2019idst}
Ming Yan, Chenliang Li, Chen Wu, Jiangnan Xia, and Wei Wang. 2019.
\newblock \href {https://trec.nist.gov/pubs/trec28/papers/IDST.DL.pdf} {Idst at
  trec 2019 deep learning track: Deep cascade ranking with generation-based
  document expansion and pre-trained language modeling}.
\newblock In \emph{Proc. of TREC}.

\bibitem[{Yang et~al.(2023)Yang, Chen, Li, Ding, and Wu}]{yang2023chatgpt}
Lin~F. Yang, Hongyang Chen, Zhao Li, Xiao Ding, and Xindong Wu. 2023.
\newblock \href {https://arxiv.org/abs/2306.11489} {{ChatGPT} is not enough:
  Enhancing large language models with knowledge graphs for fact-aware language
  modeling}.
\newblock \emph{arXiv preprint arXiv:2306.11489}.

\bibitem[{Yao et~al.(2023)Yao, Zhao, Yu, Du, Shafran, Narasimhan, and
  Cao}]{yao2023react}
Shunyu Yao, Jeffrey Zhao, Dian Yu, Nan Du, Izhak Shafran, Karthik~R Narasimhan,
  and Yuan Cao. 2023.
\newblock \href {https://openreview.net/forum?id=WE_vluYUL-X} {React:
  Synergizing reasoning and acting in language models}.
\newblock In \emph{Proc. of {ICLR}}.

\bibitem[{Zhang et~al.(2022)Zhang, Gong, Shen, Lv, Duan, and
  Chen}]{zhang2022adversarial}
Hang Zhang, Yeyun Gong, Yelong Shen, Jiancheng Lv, Nan Duan, and Weizhu Chen.
  2022.
\newblock \href {https://openreview.net/forum?id=MR7XubKUFB} {Adversarial
  retriever-ranker for dense text retrieval}.
\newblock In \emph{Proc. of ICLR}.

\bibitem[{Zhou et~al.(2023)Zhou, Shen, Geng, Tao, Xu, Long, Jiao, and
  Jiang}]{zhou2023towards}
Yucheng Zhou, Tao Shen, Xiubo Geng, Chongyang Tao, Can Xu, Guodong Long,
  Binxing Jiao, and Daxin Jiang. 2023.
\newblock \href {https://aclanthology.org/2023.findings-acl.332} {Towards
  robust ranker for text retrieval}.
\newblock In \emph{Findings of ACL}, pages 5387--5401.

\end{thebibliography}

\end{document}